# Volatile transport on inhomogeneous surfaces: I. Analytic expressions, with application to Pluto's day

Leslie A. Young

Southwest Research Institute,

1050 Walnut Street, Suite 300

Boulder, CO, 80302

`layoung@boulder.swri.edu`

## Abstract

An analytic expression for the variation in surface and sub-surface temperature is developed for worlds whose surface pressures are nearly constant with latitude and longitude and whose atmospheres are in vapor-pressure equilibrium with the dominant surface volatiles. Such worlds include the current Pluto and Triton, and other volatile-covered Kuiper Belt Objects during some portion of their heliocentric orbit. The expressions also apply on airless worlds with negligible horizontal heat flow, such as asteroids. Temperature variations in volatile-covered or bare areas as a function of time is derived in terms of three thermal parameters relating to (1) the thermal wave within the substrate, (2) the energy needed to heat an isothermal volatile slab, and (3) the buffering by the latent heat needed to change the atmospheric surface pressure. For Pluto's current surface pressure (~17 $\mu$bar), atmospheric buffering dominates over subsurface effects on diurnal timescales, and should keep the surface pressure over a Pluto day constant to within 0.2%.

# 1. Introduction

Pluto, like Triton, has an $N_2$ atmosphere in vapor-pressure equilibrium with the $N_2$ ices on its surface (Owen et al., 1993). The insolation, and the temperature of the volatile ices, varies with changing subsolar latitude and heliocentric distance. Because surface pressure is an extremely sensitive function of volatile ice temperature (Brown and Ziegler, 1980), the surface pressure is expected to vary by orders of magnitude over Pluto's and Triton's seasons (e.g., Trafton and Stern, 1983; Hansen and Paige, 1996; hereafter HP96). In fact, factor-of-two pressure variations on decadal scales have been observed on both Pluto (Elliot et al., 2003; Sicardy et al., 2003) and Triton (Elliot et al., 1998, 2000). The observed and expected changes in Pluto's atmosphere, and the planned observations of Pluto by the New Horizons spacecraft in 2015 (L. Young et al., 2008), have fueled an increased interest in Pluto's changing surface and atmosphere.

For observations of the surface, detection of secular change is complicated by Pluto's variegation, with bright areas dominated by $N_2$-rich ices, and darker areas dominated by $CH_4$ or other, involatile materials (e.g., Grundy and Fink, 1996). Because of the longitudinal variation, the change in observables over a Pluto day (6.4 Earth days) swamps that of the decadal variation, for Pluto's visible lightcurve (Buie et al., 2010), infrared spectrum (Grundy and Buie, 2001; Grundy et al., 2009), and thermal emission (Lellouch et al., 2011b).

Are similar diurnal variations complicating our interpretation of changes in Pluto's atmosphere? Pluto's atmosphere has been observed with stellar occultation at least once a year since 2006 (E. Young et al., 2008; Elliot et al., 2007; Buie et al., 2008; L. Young et al., 2009, 2010, 2011; Person et al., 2010; Olkin et al., 2012), opportunities made possible by Pluto's passage through the galactic plane (Assafin et al., 2010). The technique of stellar occultation is currently the most sensitive method for measuring the changes in Pluto's atmosphere. The high cadence of Pluto stellar occultations allows additional comparisons at the timescales of months.

At the 2010 Division of Planetary Science meeting, preliminary reports of the results of two Pluto occultations in 2010, on February 14 (Young et al., 2010) and July 4 (Person et al. 2010), suggested rapid and non-monotonic changes in Pluto's atmosphere of ~20% over 5 months, contrary to expectations of seasonal models (HP96). Two explanations were suggested at that time by Young et al. (2010). Either the differences arose from the inter-comparison of pressures derived using a variety different analysis techniques, or the differences reflected real atmospheric changes on short time scales. In particular, Young et al. (2010) noted that the pressures derived from occultations observed at a subsolar longitude of 297-12° were lower than those at subsolar longitude 82-187° (using the rotational-north-pole, sub-Charon prime meridian convention of e.g., Grundy and Fink, 1996). This paper addresses the possibility of rapid atmospheric change, particularly on diurnal timescales.

It has been argued elsewhere (Stern and Trafton, 1984; Young 1992; Stansberry et al., 1996; Spencer et al., 1997) that Pluto's surface pressure and volatile ice temperature should be nearly uniform *spatially* if the atmosphere is dense enough to effectively transport mass from areas of sublimation to areas of deposition. I ask a different, but related, question. If Pluto's atmosphere has a single surface pressure at any given time, how does that surface pressure change as Pluto rotates, so that different portions of Pluto's volatiles are illuminated? When the sun illuminates the most ice, so that the average insolation is the highest, does this lead to higher ice temperatures? In other words, is there a dependence on the subsolar longitude?

In investigating the question of Pluto's diurnal variation, I derive an analytic expression involving dimensionless thermal parameters analogous to those used by Spencer et al. (1989) for a homogeneous, volatile-free surface. The work of Spencer et al. (1989) has been widely cited, showing the utility of such thermal parameters as an aid to intuitive understanding. By including these thermal parameters in an analytical expression, their utility is extended even further, for quick investigations of the effects of different physical assumptions, and as initial conditions for numerical models.

In Section 2, I derive these thermal parameters and analytical expressions for volatile ice temperatures. I apply these parameters to the case of Pluto's diurnal variation in Section 3, and compare the expressions to previous work in Section 4.

## 2. Analytical expressions for surface and subsurface temperatures

Moore and Spencer (1990) and Spencer and Moore (1992) modeled Triton's seasonal variation with a model that included the thermal inertia of the substrate. Hansen and Paige (1992, 1996) included additionally both the thermal inertia of an isothermal volatile slab and conservation of mass between the volatile slab and the atmosphere. The conceptual framework for the model presented here is built on the physical processes considered by Hansen and Paige, 1992 and Hansen and Paige, 1996 (HP96), as illustrated in Fig. 1. These include thermal conduction into and within a substrate, an internal heat flux, absorbed sunlight, and thermal emission. For the analytic solution, the thermophysical parameters of the substrate (specific heat, $c$; thermal conductivity, $k$; and density, $\rho$) are assumed to be constant with depth and with time, but are allowed to vary from location to location. The lower boundary condition balances conduction with a specified heat flux, $F$. The emissivity, $\varepsilon$, is also taken to be variable in location but constant in time.

HP96 adopt 6 erg cm$^{-2}$ s$^{-1}$ for the internal heat flux, or about 9% of the 2011 globally averaged insolation. Note, however, that $F$ can be used to specify the lower boundary in general, and does not need to be physically identified with the internal heat flux. In particular, the seasonal thermal wave is much deeper than the diurnal thermal wave (HP96). There may be a thermal gradient at a depth of several diurnal skin depths that is a result of the seasonal thermal forcing. This can be included by using the term $F$ to account for this deep gradient, to model diurnal variation superimposed on a seasonal cycle.

**INSERT FIG 1 HERE**

This model, like that of HP96, assumes that the volatile ice temperature is the same everywhere on the planet. This assumption is discussed in detail elsewhere, (e.g., Spencer et al., 1997). In brief, if the atmosphere is dense enough to effectively transport mass from sublimation zones to condensation zones, it also transports energy in the form of latent heat. This mechanism becomes ineffective when the winds transporting the needed mass approach the sound speed (Trafton and Stern, 1983). Spencer et al. (1997) estimate that Pluto is in the global-atmosphere regime for pressures greater than 60 nbar ($N_2$ ice temperatures greater than 30.5 K). Since Pluto's atmosphere has been measured directly by stellar occultations to be higher than 7 $\mu$bar (E. Young et al., 2008), Pluto is firmly in the global-atmosphere regime.

As in HP96, this model assumes that the volatile ice[*] forms slabs that are isothermal with depth, as well as with latitude and longitude, essentially equivalent to assuming that the $N_2$ grains within the slab are in vapor pressure equilibrium with the atmosphere (Grundy and Stansberry, 2000). Within the volatile ice slab, a net energy source will lead to an increase in the volatile temperature. The $N_2$ ice in the diurnal scenario considered here does not cross the α-β phase transition of $N_2$ at 35.61 K (Brown and Zeigler, 1980), and the latent heat of the solid phase transition is thus not included in this model.

Although the latitudinally averaged volatile-transport problem has been previously implemented numerically (e.g., Moore and Spencer, 1990; HP96), there is still value in an analytic expression for the volatile transport on a longitudinally inhomogeneous surface.

---

[*] In previous work (e.g., HP96), the condensed volatile is referred to as *frost*. The term *volatile ice* or simply *volatile* is used here instead, since the term *frost* should be reserved for the condensation of a minor gaseous species diffusing through a major species (Grundy 2011).

First, simple analytic expressions provide physical insight into the dominant physical processes in a problem; it becomes easy to quantify which physical processes dominate, and which can be neglected. Second, analytic expressions provide useful diagnostics when constructing new numerical models; an estimate of expected results can be invaluable in testing code. Third, analytic expressions can provide sensible initial conditions for numerical calculations.

To make this problem amenable to an analytic solution, elements on the surface are described as either volatile-covered or volatile-free. This is a simplification for two reasons. The first is that the surface of Pluto consists of the volatile species CO and $CH_4$ as well as $N_2$. A volatile-transport model that includes these minor species will be an important extension, especially for understanding the evolution of the atmospheric composition of volatile-covered bodies such as Pluto and Triton. However, while multi-component volatile ices can radically alter the surface pressure in some cases, through the formation of a $CH_4$ or CO rich crust (Trafton 1990), it appears as though any such crust on Pluto or Triton, if present, is not rich enough in the minor species to shut off the communication of $N_2$ with the atmosphere (Lellouch et al., 2009, 2011a). Therefore, the usual assumption (HP96, Spencer et al., 1997), that the atmospheric $N_2$ is in vapor-pressure equilibrium with the $N_2$ ice, appears valid. Therefore, for application to Pluto or Triton, "volatile-covered" refers to areas of $N_2$-dominated ices, while "volatile-free" refers to both $CH_4$-dominated areas, and areas devoid of volatiles. The second reason why describing an area on the surface as either volatile-covered or volatile-free (that is, a static composition model) is a simplification is that, in reality, volatiles are free to move around the surface. For Pluto's diurnal cycle, the net sublimation or deposition (~10 micron) is a small fraction of the $N_2$ grain size (Grundy and Stansberry, 2000), justifying this static description of the volatile distribution. A static distribution is less valid at decadal timescales, as $N_2$ is transported from areas of net sublimation to net deposition. For example, in run #12 of HP96, areas of the sunlit pole become volatile-free at a rate of 1% of Pluto's total surface area per year in the current post-perihelion epoch. Over seasonal timescales, a static distribution is only valid if a surface is an "ice ball" that is entirely covered in volatiles, or if high substrate thermal inertia leads to a static volatile distribution (e.g., the "Koyaanismuuyaw" model of hemispherical dichotomy of Moore and Spencer, 1990, or run # 32 of HP96).

Given solar forcing that repeats over a period $P$ (e.g., a rotation period or a year), we are looking for periodic solutions of the surface temperatures. The simplest such solution is one where the absorbed insolation, $S$, and volatile temperature, $T_V$, are expressed as the sum of a constant term and a sinusoidal term. This is most compactly expressed in its complex form:

$$S(t,\lambda,\phi) = \hat{S}_0(\lambda,\phi) + \hat{S}_1(\lambda,\phi)e^{i\omega t}$$
$$T_V(t) = \hat{T}_{V0} + \hat{T}_{V1}e^{i\omega t} \tag{1}$$

where $t$ is time, $\omega = 2\pi/P$ is the frequency, $\lambda$ is latitude, and $\phi$ is longitude. The mean quantities, $\hat{S}_0$ and $\hat{T}_{V0}$, are real. The coefficients of the sinusoidal portion, $\hat{S}_1$ and $\hat{T}_{V1}$, are complex. As usual for the complex representation of wave equations, the real part is taken for the physical quantities. If $\hat{S}_1$ and $\hat{T}_{V1}$, are expressed in terms of real amplitudes and phases (with the sign of the phase chosen so positive phases indicate a temporal lag), $\hat{S}_1 = \left|\hat{S}_1\right|\exp(-i\psi_S)$ and $\hat{T}_{V1} = \left|\hat{T}_{V1}\right|\exp(-i\psi_T)$, then the solar forcing and thermal response can be expressed entirely with real numbers: $S = \hat{S}_0 + \left|\hat{S}_1\right|\cos(\omega t - \psi_S)$ and $T_V = \hat{T}_{V0} + \left|\hat{T}_{V1}\right|\cos(\omega t - \psi_T)$. The main goal of this paper is to derive the relationship between the solar and temperature amplitudes, $\left|\hat{T}_{V1}\right|/\left|\hat{S}_1\right|$, and the lag between the phases, $\psi = \psi_T - \psi_S$.

At latitudes where the sun never sets (or, trivially, where it never rises), the insolation is exactly described by a sinusoid. Elsewhere, the insolation, like any arbitrary function, can be described as a sum of Fourier terms: $S(t,\lambda,\phi) \approx \hat{S}_0 + \hat{S}_1\exp(i\omega t) + \hat{S}_2\exp(2i\omega t) + \cdots$. For example, for the sun at the equator, $\hat{S}_0 = (S_{1AU}/\Delta^2)(1-A)/\pi$, $\hat{S}_1 = (\pi/2)\hat{S}_0$, $\hat{S}_2 = (2/3)\hat{S}_0$, $\hat{S}_3 = 0$, $\hat{S}_4 = -(2/15)\hat{S}_0$, etc., where $S_{1AU}$ is the normal insolation at 1 astronomical unit (AU), $\Delta$ is the heliocentric distance in AU, and $A$ is the wavelength-averaged hemispheric albedo (the local equivalent of Bond albedo, Hapke 1993). Higher order frequencies can be included in the expansion of insolation and temperature in Eq, (1), with which the true insolation (seasonal or diurnal) can be more accurately estimated, without the need for a complete numerical implementation.

The temperature within the substrate, $T_S$, satisfies the diffusion equation, $\rho c \dot{T}_S = k \, \partial^2 T_S / \partial z^2$ (HP96), where dotted variables indicate derivatives with respect to time, and $z$ is the height above the substrate surface (zero at the top of the substrate, decreasing downward). At the boundary between the substrate and the volatile slab, the substrate temperature equals the slab temperature. At the lower boundary, $k \, \partial T_S / \partial z = -F$. A temperature profile of the form

$$T_S(z,t) = -(F/k)z + \hat{T}_{V0} + \hat{T}_{V1} e^{i\omega t + (\sqrt{i\omega}\,\Gamma/k)z} \tag{2}$$

satisfies the diffusion equation and the boundary conditions, where $\Gamma = \sqrt{k\rho c}$ is the thermal inertia, introduced to simplify later expressions. This can be confirmed with direct substitution into the diffusion equation and boundary condition, noting that, if $w \equiv \exp[i\omega t + (\sqrt{i\omega}\,\Gamma/k)z]$, then $\dot{T}_S = i\omega \hat{T}_{V1} w$, $\partial^2 T_S / \partial z^2 = i\omega (\Gamma^2/k^2) \hat{T}_{V1} w$, and $\partial T_S / \partial z \rightarrow -(F/k)$ for large negative values of $z$. The skin depth, $Z$, as defined by Spencer et al. (1989) and HP96, is $Z = k/(\Gamma \omega^{1/2})$. Since $\sqrt{i} = (1+i)/\sqrt{2}$, the time-variable term describes a damped oscillation, with wavelength $2\pi\sqrt{2}Z$ and e-folding distance of $\sqrt{2}Z$.

As illustrated in Fig. 1, heating of the volatile slab depends on solar insolation (always a source), thermal emission (always a sink), thermal conduction to the substrate (a sink if the volatile slab is warmer than the substrate), and latent heat (positive for deposition, or $\dot{m}_V > 0$).

$$c_V m_V \dot{T}_V = S - \varepsilon \sigma T_V^4 - k \frac{\partial T_S}{\partial z}\bigg|_{z=0} + L\dot{m}_V \tag{3}$$

where $c_V$ is the specific heat of the volatile slab (subscripted $V$ for volatile), $m_V$ is the column mass of the volatile slab, $\varepsilon$ is the thermal emissivity, $\sigma$ is Stefan-Boltzmann constant, $L$ is the latent heat of sublimation, and $S$ is the absorbed insolation, given by $S = \left(S_{1AU}/\Delta^2\right)\cos\theta(1-A)$, where $\theta$ is the incidence angle. This expression ignores the effects of surface roughness on the surface temperature (Spencer 1990). The temperature derivative is calculated at top of the substrate, at $z = 0$.

The spatial average over the volatile-covered areas is denoted by angled brackets. For example, the insolation averaged over the volatiles is

$$\langle S(t)\rangle \equiv \frac{1}{4\pi f_V} \int\limits_{volatile} S(t,\lambda,\phi)\cos\lambda\, d\lambda\, d\phi \tag{4}$$

where $f_V$ is the fractional area covered by volatiles.

Global energy balance is then found by averaging the local energy balance (Eq. 3) over the volatiles:

$$\langle c_V m_V\rangle\dot{T}_V = \langle S\rangle - \langle\varepsilon\rangle\sigma T_V^4 - \left\langle k\frac{\partial T_S}{\partial z}\bigg|_{z=0}\right\rangle + L\langle\dot{m}_V\rangle \tag{5}$$

where we have used the fact that $T_V$ is constant over all the areas covered by volatiles to factor it outside the angled brackets.

To proceed further, we need to eliminate the final term, $L\langle\dot{m}_V\rangle$. This is sometimes achieved by imposing $\dot{m}_V = 0$, or global balance of sublimation and deposition (e.g., Moore and Spencer 1990). It is more correct to consider global mass balance, including the change in the atmospheric bulk (e.g., HP96). The equation for global mass balance is

$$f_V\langle\dot{m}_V\rangle + \dot{m}_A + E = 0 \tag{6}$$

where $m_A$ is the column mass of the atmosphere (assumed to be globally uniform), and $E$ represents the globally averaged escape rate, in units of mass per area per time. As with the volatile temperature and solar forcing, the escape rate is expressed as a sinusoid, as $E = \hat{E}_0 + \hat{E}_1 e^{i\omega t}$, where $\hat{E}_1$ is a complex number. The motivation for including $E$ is two-fold. One is to allow comparisons in later papers with the models such as Trafton (1990), where the seasonally variable escape rate is a critical factor in describing the seasonal evolution. The other is to allow the inclusion of sources other than sublimation from the volatile slab (which show up as a negative contribution to $E$). This includes geysers, such as on Triton, or as an ad-hoc method of including sublimation of $N_2$ from nearly pure $CH_4$ hotspots (Stansberry et al., 1996) until a two-species model can be constructed.

The column mass of the atmosphere is very nearly a function only of the surface temperature (ignoring a small correction term, proportional to the ratio of the atmospheric scale height to the surface radius). This means that the change in atmospheric mass can be expressed as

$$\dot{m}_A = \frac{dm_A}{dT_V} \dot{T}_V \tag{7}$$

Hydrostatic equilibrium relates $m_A$ and the surface pressure, $p_{surf}$, through the effective gravitational acceleration, defined by $g = p_{surf} / m_A$. For bodies with large scale heights, $H$, such as Pluto, this includes an adjustment to the gravitational acceleration at the surface, $g_{surf}$, so that $g = g_{surf}(1 - 2H/R)$. The Classius-Clapeyron relation states that $dp_{surf}/dT_V = L_T p_{surf}/T_V^2$, where $L_T$ is latent heat expressed in units of temperature ($L_T = L\mu m_{amu}/k_B$, where $\mu$ is the molecular weight, $m_{amu}$ is the atomic mass unit, and $k_B$ is Boltzmann constant. For $N_2$ at 37.9 K, $L_T = 852.7$ K). $L_T$ is introduced purely for notational convenience. Using hydrostatic equilibrium and integrating the Classius-Clapeyron relation gives

$$\frac{dm_A}{dT_V} = \frac{1}{g}\frac{dp_{surf}}{dT_V} = \frac{1}{g}\frac{L_T p_0}{T_V^2}\exp\left[L_T\left(\frac{1}{T_{V0}} - \frac{1}{T_V}\right)\right] \tag{8}$$

where $p_0$ is the equilibrium vapor pressure at $T_{V0}$.

Substituting Eqs. 6 and 7 into Eq. 5 gives a new form of the global energy equation:

$$\left(\langle c_V m_V \rangle + \frac{L}{f_V}\frac{dm_A}{dT_V}\right)\dot{T}_V = \langle S \rangle - \langle \varepsilon \rangle \sigma T_V^4 - \left\langle k\frac{\partial T}{\partial z}\bigg|_{z=0}\right\rangle - \frac{L}{f_V}(E) \tag{9}$$

We substitute the sinusoidal expressions for $T_V$ and $S$ (Eq. 1) into Eq. (9) and expand to first order. The time-averaged form of Eq. (9) becomes

$$0 = \langle \hat{S}_0 \rangle - \langle \varepsilon \rangle \sigma \hat{T}_{V0}^4 + \langle F \rangle - \frac{L}{f_V}\hat{E}_0 \tag{10}$$

which simply states that the thermal emission balances solar insolation, internal heat flux, and the latent heat associated with atmospheric sources and sinks. By restricting the expansion to first order, we are ignoring cross terms in $T_V^4$ that depress the mean temperature in the case of small thermal inertia, as seen in Spencer et al. (1989). This is discussed further in Section 3.

The time-varying portion of Eq. (9), or those terms proportional to $e^{i\omega t}$, are

$$\left(\langle \hat{m}_{V0} c_V \rangle + \frac{L}{f_V}\frac{dm_A}{dT_V}\right)i\omega\hat{T}_{V1} = \langle \hat{S}_1 \rangle - 4\langle \varepsilon \rangle \sigma \hat{T}_{V0}^3 \hat{T}_{V1} - \sqrt{i\omega}\langle \Gamma \rangle \hat{T}_{V1} - \frac{L}{f_V}\hat{E}_1 \tag{11}$$

where $\hat{m}_{V0}$ is mean mass of the volatile slab.

Eq. (11) is written more simply by defining three thermal parameters (Eqn. 12-14). The first, $\Theta_S$, describes the buffering of the modulation of the volatile ice temperature due to thermal conduction into and out of the substrate (subscripted $S$ for substrate). This parameter is a generalization of the thermal parameter defined in Spencer et al. (1989). Where Spencer et al. (1989) defined their thermal parameter $\Theta$ relative to the sub-solar equilibrium temperature, $T_{SS}$, in this work $\Theta_S$ is a function of temperature:

$$\Theta_S(T) = \frac{\sqrt{\omega}\langle \Gamma \rangle}{\langle \varepsilon \rangle \sigma T^3} \tag{12}$$

The second thermal parameter, $\Theta_V$, describes the buffering of the modulation of the volatile temperature due to the thermal inertia of the isothermal volatile ice slab (subscripted $V$ for volatile slab).

$$\Theta_V(T) = \frac{\omega \langle \hat{m}_{V0} c_V \rangle}{\langle \varepsilon \rangle \sigma T^3} \tag{13}$$

The final parameter, $\Theta_A$, describes how the atmosphere buffers the volatile ice temperature due to the latent heat needed to change the atmospheric pressure in response to the volatile ice temperature (subscripted $A$ for atmosphere).

$$\Theta_A(T) = \omega \frac{L}{f_V} \frac{dm_A(T)}{dT_V} \frac{1}{\langle \varepsilon \rangle \sigma T^3} \tag{14}$$

With these three parameters, the variation in the temperature can be written to first order as

$$\hat{T}_{V1} = \frac{\langle \hat{S}_1 \rangle - \frac{L}{f} \hat{E}_1}{4 \langle \varepsilon \rangle \sigma \hat{T}_{V0}^3} \frac{4}{4 + \sqrt{i} \Theta_S(\hat{T}_{V0}) + i\left(\Theta_V(\hat{T}_{V0}) + \Theta_A(\hat{T}_{V0})\right)} \tag{15}$$

A similar derivation applies to volatile-free regions, where the mass of the volatile slab is zero and there is no latent heat term in the local energy balance equation (Eq. 3). In this case, there is no communication between the volatile-free surface elements. Writing the local bare (i.e., volatile-free) surface temperature as $\hat{T}(t,\lambda,\phi) = \hat{T}_{B0}(\lambda,\phi) + \hat{T}_{B1}(\lambda,\phi)e^{i\omega t}$, the volatile-free equation for the time-averaged temperature (equivalent to Eq. 10) is

$$0 = \hat{S}_0 - \varepsilon \sigma \hat{T}_{B0}^4 + F \tag{16}$$

and the volatile-free variation is, analogous to Eq. (15),

$$\hat{T}_{B1} = \frac{\hat{S}_1}{4 \varepsilon \sigma \hat{T}_{B0}^3} \frac{4}{4 + \sqrt{i} \Theta_S(\hat{T}_{B0})} \tag{17}$$

where the thermal parameter $\Theta_S$ is calculated using local values of $\Gamma$, $\varepsilon$, and $T_{B0}$, or $\Theta_S(T) = \sqrt{\omega} \Gamma / (\varepsilon \sigma T^3)$.

For languages with that support it, it is simplest to calculate Eqs. (15) and (17) in complex arithmetic, and then take the real part of the complex expression $T_V(t) = \hat{T}_{V0} + \hat{T}_{V1} e^{i\omega t}$ or $T(t,\lambda,\phi) = \hat{T}_{B0}(\lambda,\phi) + \hat{T}_{B1}(\lambda,\phi)e^{i\omega t}$ as the physical quantity. This can be described working strictly in real arithmetic as well. As discussed in the text following Eq. (1), $\hat{S}_1$ is a complex number with amplitude $\left|\hat{S}_1\right|$ and phase $-\psi_S$, giving rise to a time-dependant term $\left|\hat{S}_1\right|\cos(\omega t - \psi_S)$. The net forcing in Eq. (15), $\langle\hat{S}_1\rangle - (L/f)\hat{E}_1$, is also complex, and can be similarly written with an amplitude and phase. The denominator, $4\langle\varepsilon\rangle\sigma\hat{T}_{V0}^3$ or $4\varepsilon\sigma T_{B0}^3$, is a real number. Therefore, if the thermal parameters are all zero, the temperature will track the forcing, with no change in phase, giving $\hat{T}_{V1,eq} = [\langle\hat{S}_1\rangle - (L/f)\hat{E}_1]/(4\varepsilon\sigma T_0^3)$ or $\hat{T}_{B1,eq} = \hat{S}_1/(4\varepsilon\sigma T_{B0}^3)$. Therefore the first multiplicative term in Eq. (15) or (17) represents the variation in the equilibrium temperature, $\hat{T}_{V1,eq}$ or $\hat{T}_{B1,eq}$, or the time-dependant portion of the volatile temperature in instantaneous balance with the insolation, internal flux, and escape and other mass source terms.

The three thermal parameters derived here allow a quick intuitive insight into the effect of thermal inertia and atmospheric buffering. As described by Spencer et al. (1989), small values of the thermal parameters leads to temperatures that strongly track the solar input with only small phase lags, while large thermal parameters strongly suppress the temperature variation and leads to a phase lag that tends to an asymptotic value (Table I). The amplitude of the temperature variation is suppressed by a factor $a$ (a real number):

$$a = \left|\frac{T_{V1}}{T_{V1,eq}}\right| = \frac{4}{\sqrt{\left(4 + \Theta_S/\sqrt{2}\right)^2 + \left(\Theta_V + \Theta_A + \Theta_S/\sqrt{2}\right)^2}} \tag{18}$$

For small values of $\Theta_S$, $\Theta_V$, and $\Theta_A$, $a \approx 1$, and the temperature approximates the equilibrium temperature. If any of the thermal parameters are large, then $a \ll 1$, and the temperature variation is suppressed. This is quantified in Table I, which gives expressions for $a$ for cases considering only one thermal parameter at time, for the limits of large or small thermal parameters.

The phase lag, $\psi$, a real number, can be derived from the real and imaginary parts of the denominator in Eq. (15), recalling that $\sqrt{i} = (1+i)/\sqrt{2}$, so that

$$\psi = \arctan\left(\frac{\Theta_V + \Theta_A + \Theta_S/\sqrt{2}}{4 + \Theta_S/\sqrt{2}}\right) \qquad (19)$$

The phase lag is small (Table I) if the thermal parameters are small, and asymptotes to $\pi/4$ for large values of $\Theta_S$, and $\pi/2$ for large values of $\Theta_V$ or $\Theta_A$.

The relaxation timescale is the characteristic timescale for the temperature to relax to a new equilibrium. The derivation for the timescale proceeds analogously as above, expressing the temperature as $T_V(t) = T_{V0} + T_{V1}e^{-t/\tau}$. The timescale is then easily expressed in terms of $\Theta_V$, $\Theta_A$ and $\Theta_S$.

$$\tau = \frac{1}{\omega}\left(\frac{\Theta_V}{4} + \frac{\Theta_A}{4} + \frac{\Theta_S}{4\sqrt{2}}\right) \qquad (20)$$

## INSERT TABLE I HERE

Eqs. (10) and (15) are used to find the variation of volatile temperature with time. I emphasize that this formulation describes the variation of the volatile temperature with *time*, while maintaining usual assumption that the volatile temperature is *spatially* isothermal.

Returning to the local energy equation, Eq. (3), we can derive the equation for the local sublimation rate ($\dot{m}_V = \dot{m}_{V0} + \dot{m}_{V1}e^{i\omega t}$) averaged over a period. This is simply the local energy input minus the energy input averaged over the volatiles.

$$-L\dot{m}_{V0} = \left(\hat{S}_0 + F\right) - \left(\langle\hat{S}_0\rangle + \langle F\rangle - \frac{\hat{E}_0}{f}\right) \qquad (21)$$

The time-varying portion of the sublimation rate is similarly related to the time-varying difference between the local insolation and that averaged over the volatile areas.

$$-L\dot{m}_{V1} = \hat{S}_1 - \langle\hat{S}_1\rangle + \frac{i\Theta_A}{4 + \sqrt{i}\Theta_S + i\Theta_V + i\Theta_A}\langle\hat{S}_1\rangle \qquad (22)$$

If the time-varying average insolation, $\left\langle \hat{S}_1 \right\rangle$, is small, or if $\Theta_A$ is large, then the local sublimation rate balances the variation of the local insolation. If $\Theta_S$ or $\Theta_V$ are large and $\Theta_A$ is small, so that temperatures are buffered by something other than the atmosphere, then the local sublimation rate balances the difference between local and global insolation.

## 3. Application to Pluto

This work was motivated by the question of whether or not we expect to see temporal variation in Pluto's $N_2$ ice temperature and surface pressure, over the course of a Pluto day. I repeat that this is different than the spatial variation of $N_2$ ice temperature and surface pressure over Pluto's surface, which has been treated elsewhere (Stern and Trafton, 1984; Young 1992; Stansberry et al., 1996; Spencer et al., 1997). To address this question, I consider the constraints on Pluto's thermal inertia derived from its thermal emission by Lellouch et al. (2011b; hereafter L11). L11 considered several composition maps, including the map of Grundy and Fink (1996). For this map, L11 use $A = 0.67$, $\varepsilon = 0.5$, and $T_V = 37.4$. These are not self-consistent; that is, the insolation averaged over the $N_2$ ice, $\langle S \rangle$, is 107 erg cm$^{-2}$ s$^{-1}$, while the thermal emission, $\varepsilon \sigma T_V^4$, is only 78 erg cm$^{-2}$ s$^{-1}$. This inconsistency has little impact on the modeling or conclusions of L11, since they assume a $N_2$ ice temperature that is constant with time and location. Because I am modeling the time variation of the $N_2$ ice temperature, I chose similar, but self-consistent, parameters: $A = 0.8$, $\varepsilon = 0.55$, and $T_V = 37.87$ K. The result, that Pluto's atmosphere is constant over a day to better than 1%, is robust to details of the choices of albedo, emissivity, and heat flux, as described below.

**INSERT FIG 2 HERE**

In the Grundy and Fink (1996) composition map, the $N_2$-rich terrain covers 53% of the total area of Pluto, but is unevenly distributed (Fig 2A). Assuming all the $N_2$ ice has the same albedo, then disk-averaged insolation absorbed by the volatile ice is proportional to its projected area. As seen from the Sun in 2011, the $N_2$-rich areas cover 65% of the visible disk at a longitude of 209°, near lightcurve maximum (Buie et al., 2010), but only 41% at a longitude of 77°, near lightcurve minimum (Fig 2B). Therefore, the insolation is a factor of 1.6 times higher for a sub-solar latitude of 209° than 77° (81 vs. 52 erg cm$^{-2}$ s$^{-1}$). If the N$_2$ ice temperature were such that the thermal emission was in instantaneous equilibrium with the insolation averaged over the volatiles (e.g., $\varepsilon \sigma T_{V,eq}^4 = \langle S(t) \rangle$), then the N$_2$ ice temperatures would range from 36.0 to 40.2 K. The pressure is a sensitive function of temperature, and this 4.2 K peak-to-peak variation (Fig 2C, solid) implies that the surface pressure would vary by a factor of 12 (Table II). The sinusoidal approximation for the solar forcing, as a function of sub-solar longitude $\phi_0$, is $\langle S(t) \rangle = 64.14 + 13.34 \cos(\phi_0\text{-}216°)$ erg cm$^{-2}$ s$^{-1}$; for the temperature it is $T_{V,eq} = 37.9 + 2.0 \cos(\phi_0\text{-}216°)$ K (Fig 2C, dashed), giving a similar pressure variation (pressures varying by a factor of 10). This huge range of pressure variation is not seen in the history of Pluto occultations (e.g., Young et al., 2010, Person et al., 2010).

## INSERT TABLE II HERE

I next consider the thermal inertia of the substrate. I adopt a thermal inertia of 1.8x10$^4$ erg cm$^{-2}$ K$^{-1}$ s$^{-1/2}$ based on Lellouch et al., 2011b, using their $\Theta_{PL} = 5$ from the Grundy and Fink (1996) map (line 2 of L11 Table 4). This gives a diurnal thermal parameter for the substrate, $\Theta_S$, of 36. This is larger than the values of $\Theta(T_{SS}) \sim 5\text{-}12$ in L11 for the CH$_4$-rich and tholin areas because that is calculated for the equilibrium sub-solar temperature of 63.3 K, rather than for $T_{V0} = 37.9$ K. The thermal inertia decreases the temperature variation of the N$_2$ ice by roughly a factor of 9 (or $\Theta_S/4$), for a 0.4 K peak-to-peak range (Fig 2D). While much smaller than the variation for equilibrium temperatures, this is still large enough for the surface pressure to vary by 27%. This is precisely the size of variation that motivated this work. However, the two other effects, the thermal inertia of the volatile slab and the atmospheric buffering, decrease this variation still further.

An isothermal N$_2$-ice slab, with $c_V = 1.3$x10$^7$ erg g$^{-1}$ K$^{-1}$ (Spencer and Moore, 1992), and m$_V = 9.5$ g cm$^{-2}$, based on run #12 of HP96, implies a diurnal thermal parameter for the N$_2$ ice slab, $\Theta_V$, of 830. This restricts the temperature range over a day to 0.012 K peak-to-peak (Fig 2E), and the pressure variation to only 1.1%. If subsurface processes were the only processes applicable, then we could constrain the depth to which the solid-gas energy exchange keeps the $N_2$ slab isothermal by observations of the diurnal variation of pressure. However, as I now show, the atmospheric buffering overwhelms the buffering due to the thermal inertia of the substrate or volatile slab.

For the adopted albedo and emissivity, the temperature is 37.87 K, and the surface pressure is 17.4 $\mu$bar, similar to the surface pressure derived by Lellouch et al., (2009). In the Grundy and Fink (1996) composition map, the fraction of Pluto's surface covered by volatiles, $f_V$, is 53% (by surface area, not projected area). Using Eq. (14), the thermal parameter for the atmosphere, $\Theta_A$, is 5411. The atmospheric buffering restricts the change in the temperature of the $N_2$ ice within the volatile slab over a Pluto day to only 0.002 K (Fig 2F), and the variation in pressure to only 0.15%.

This work implies that if variation is seen in Pluto's atmosphere on short time scales, then the change must be due to something other than changes in surface pressure. One such physical change that can affect a lightcurve is variation in the thermal structure due variation in the atmospheric $CH_4$ or CO mixing ratio.

## 4. Comparison with previous work

In analyzing thermal emission from Pluto, Lellouch et al. (2000, 2011b) assume that the $N_2$ ice temperature is independent of location and time. This work affirms the validity of that assumption.

The formulation of the thermal parameters was constructed in parallel to the work of Spencer et al. (1989), who derived a thermal parameter based on thermal inertia and the equilibrium sub-solar temperature, with a form similar to $\Theta_S$ (compare Eq. (12) in this work and Eq. (7) in Spencer et al., 1989). The equations in this paper are most appropriate when the solar forcing is sinusoidal and the temperature variation is small. Even in cases where this is not the case, such as on the equator of an airless body (Spencer et al. 1989), this work gives useful approximations. For an equatorial surface element at equinox, the normal insolation is larger than the diurnally averaged insolation by a factor of $\pi$, so $T_{SS}$ will be larger than $\hat{T}_{B0}$ by a factor of $\pi^{1/4}$. Correspondingly, $\Theta_S(T_{ss})$, the thermal parameter from Spencer et al. (1989), will be smaller than $\Theta_S(T_{B0})$, the thermal parameter used here, by $\pi^{3/4} = 2.4$. Fig. 3 shows the surface temperature for an equatorial element at equinox, following Spencer et al., (1989). For all values of $\Theta_S(T_{ss})$, the analytic expressions for a volatile-free area (Eq. 16 and 17) reproduce the phase shift and amplitude of the temperature variation well. For example, for $\Theta_S(T_{ss}) = 1$ ($\Theta_S(\hat{T}_{B0}) = 2.4$), Eq. (19) predicts a lag of 17°, while numerical integration predicts a lag of 11°. For $\Theta_S(T_{ss}) = 10$ ($\Theta_S(\hat{T}_{B0}) = 24$), Eq. (19) predicts a lag of 39°, while numerical integration predicts a lag of 33°.

However, the analytic expressions only reproduce the mean temperatures well for $\Theta_S(T_{ss}) > 1.0$ ($\Theta_S(\hat{T}_{B0}) > 2.4$). This is because the true mean temperature decreases with decreasing $\Theta_S$, a second-order effect not captured in this first-order analysis. While the analytic equations could be extended to include the second-order cross terms (the effect of the temperature variation on the mean temperature), the resulting equations are complex enough to offer no advantage over numerical integration of the diffusion equation. This is especially true in light of the efficacy of the analytic solution to establish an improved initial condition for numerical integration. Even with $\Theta_S(T_{ss}) = 0.1$, a numeric integration that begins with the analytic solution as the initial condition converges to the exact numeric solution by 1/10 to 1/4 of a period.

**INSERT FIG 3 HERE**

Trafton and Stern (1983) considered how long it would take half the atmosphere to freeze out if the insolation were instantaneously "turned off," and derive a timescale of

$$\tau_A = m_A L / 2\varepsilon\sigma T_V^4.$$ They present a timescale of 93 Earth days for $CH_4$ at $T_V = 57.8$ K, much

longer than the 3.2 Earth days of night for Pluto at equinox, and conclude that there is little freezeout overnight. For $N_2$, with $m_A$=0.28 g cm$^{-2}$, this timescale is smaller, 65 Earth days, but still much longer than 3.2 Earth days. The scenario of insolation instantaneously "turning off" is applicable on a seasonal timescale to the case of the disappearance by sublimation of a summer cap, leaving an unilluminated winter cap (HP96). The timescale derived here, 1375 days, is longer because $dm_A/dT_V >> m_A/T_V$.

$$\tau_A = L \frac{m_A}{T_V} \frac{1}{2\varepsilon\sigma T_V^3} \qquad \text{Turning off of insolation, Trafton \& Stern, 1983}$$

$$\tau_A = \frac{L}{f_V} \frac{dm_A}{dT_V} \frac{1}{4\varepsilon\sigma T_V^3} \qquad \text{Relaxation to new equilibrium, this work}$$

(23)

Stansberry et al. (1996) introduce a measure of the relative importance of latent heat relative to thermal inertia, which they call $I$, since it is a measure of the isothermality:

$$I = \frac{L}{\varepsilon\sigma T^4} \dot{m}_V = \frac{L}{\varepsilon\sigma T^4} \frac{p\Pi}{c_S}$$

(24)

where $c_S$ is the speed of sound, and $\Pi$ is the is the fractional difference between the equilibrium vapor pressure and the actual surface pressure. Stansberry et al. (1996) estimate the relative importance of latent heat in quasi-steady state (e.g., the ratio of the energy fluxes for latent heat and thermal emission), while this work estimates the relative importance of latent heat in response to an external forcing (e.g., the ratio of the *temperature derivative* of energy fluxes of latent heat or thermal emission). However, it is interesting to note that $\Theta_A = L\dot{m}_V / (f_V\varepsilon\sigma T^4)$, so $I$ and $\Theta_A$ should be comparable. In fact, Stansberry et al. (1996) derive $I = 5200$, remarkably similar to the diurnal value of $\Theta_A = 5411$ from Table II.

While we can use the above analysis to quantify Pluto's behavior on diurnal timescales, it should be used with caution for seasonal evolution. One of the simplifying assumptions is that the distribution of volatile ices is static, something that is certainly not the case with Pluto (HP96). However, we can get a rough feel for the relative importance of the various buffering mechanisms by considering an "ice ball," or a world covered entirely with volatiles (Table III).

**INSERT TABLE III HERE**

The first row in Table III shows the large variability in Pluto's atmosphere expected from its large range in heliocentric distances. As reviewed in Spencer et al. (1997), between perihelion at 29.7 AU and aphelion at 49.1 AU, insolation varies by a factor of 2.8; for $A = 0.8$, this range is 77.7 to 28.3 erg cm$^{-2}$ s$^{-1}$. For an unbuffered iceball with 6 erg cm$^{-2}$ s$^{-1}$ of internal heat flux (HP96) and $\varepsilon = 0.55$, the temperature would vary from 40.5 K at perihelion to 32.4 K at aphelion, with the pressure varying by a factor of 213 (74 to 0.35 $\mu$bar). The sinusoidal approximation (as a function of the mean anomaly, $M$) gives similar results, with $<S> = 45.42 + 22.32 \cos(M)$ erg cm$^{-2}$ s$^{-1}$, $T_{V,eq} = 35.8 + 3.9 \cos(M)$ K, and a pressure variation of a factor of 208. Other choices of $A$ and $\varepsilon$ give similarly large pressure swings, many of which would lead to atmospheres that drop below the critical threshold for maintaining a global atmosphere. For example, for $\varepsilon = 1$, the pressure varies from 2.4 $\mu$bar at perihelion to 0.004 $\mu$bar at aphelion.

Although Lellouch et al. (2011b) derived a low thermal inertia for diurnal variation, it is likely that the substrate becomes more compact with depth. A seasonal thermal inertia that is larger than the diurnal is consistent with the comparison of the HP96 models with stellar occultations. The low and moderate thermal inertia runs of HP96 use $4 \times 10^4$ and $2.9 \times 10^5$ erg cm$^{-2}$ K$^{-1}$ s$^{-1/2}$ respectively; models with both low and moderate inertia reproduce the rise in surface pressure between 1988 and 2006 (E. Young et al. 2008; Elliot et al. 2007), but more recent occultations 2007-2010 favor moderate thermal inertia (Young et al., 2010). As seen in the second row of Table III, including a moderate thermal inertia has a profound effect on the range of temperatures and pressures, reducing the range of pressures to only a factor of ~10. Pluto is *not* an iceball, but if it were, it is likely that the thermal inertia would keep the ice temperatures in a range that can support a global atmosphere throughout the year. The implication for other KBOs is that thermal inertia and the seasonal history of the surface should be included when interpreting thermal emission.

The third row of Table III shows that, for this example, the specific heat of the volatile slab is not important in seasonal models, compared to the thermal inertia of the substrate. For smaller thermal inertia, the roles would be reversed.

The fourth row of Table III shows that the atmospheric buffering has a small effect on seasonal timescales. This is mainly because $\Theta_A$ is inversely proportional to the period, which is about $1.4 \times 10^4$ times longer for Pluto's season than for Pluto's day. Additionally, $dp/dT$ is smaller than for the diurnal example, because of the colder mean ice temperature. While short on the seasonal timescale, the time constant of 0.7 years comparable to the annual Pluto observing season.

## 5. Summary and conclusions

I derive an analytic expression for the variation in surface and sub-surface temperature on a world where the pressure is in vapor-pressure equilibrium with surface volatile ices. This is most simply described in terms of three thermal parameters relating to the thermal wave within the substrate, the energy needed to heat an isothermal slab of volatile ice, and the buffering by the latent heat needed to change the atmospheric pressure. One of these thermal parameters is identical in form to that derived by Spencer et al. (1989), and another is closely related to the time constant for thermal collapse derived by Trafton and Stern (1983). Nevertheless, this is the first application of these thermal parameters for calculating an approximate temperature field (temperature as a function of latitude, longitude, depth, and time) for volatile-covered or volatile-free areas on a Pluto-like body.

This approximation can be used as an aid to intuition. If any one of the three thermal parameters is large, or the periodic variation in the solar forcing is small, this approximation can be used for rapid computation of the temperature field. Even for smaller values of the thermal parameters, where the agreement with numerical calculations is less accurate, the approximation can be used as an effective initial condition, allowing convergence to the final solution in typically 1/10 to 1/4 of a period.

The analytic model was applied to an example describing Pluto's diurnal variation. For Pluto's current surface pressure (~17 $\mu$bar), atmospheric buffering dominates over subsurface effects, and should keep the surface pressure over a Pluto day constant to within 0.2%. Any short-term variations in observations of Pluto's atmosphere are the product of changes in the temperature profile or atmospheric composition, not changes in the surface pressure. Such observations include high-resolution infrared absorption lines or the shape of occultation lightcurves.

## Acknowledgement

This work was supported, in part, by funding from NASA Planetary Atmospheres grant NNG06GF32G and the Spitzer project (JPL research support agreement 1368573). This paper was improved by discussions with or critical readings by Marc Buie, Will Grundy, John Spencer, and two anonymous referees.

## Tables

### Table I. Thermal parameters and their effect on the temperature response

| | Thermal parameter $\Theta$ | Amplitude factor, $a$ | | Phase lag (radian), $\psi$ | | Relaxation time scale, $\tau$ |
|---|---|---|---|---|---|---|
| | | $\Theta \ll 4$ | $\Theta \gg 4$ | $\Theta \ll 4$ | $\Theta \gg 4$ | |
| Substrate | $\dfrac{\sqrt{\omega}\langle\Gamma\rangle}{\langle\varepsilon\rangle\sigma T_{V0}^3}$ | $1-\dfrac{\Theta_S}{4\sqrt{2}}$ | $\dfrac{4}{\Theta_S}\left(1-\dfrac{2\sqrt{2}}{\Theta_S}\right)$ | $\dfrac{\Theta_S}{4\sqrt{2}}$ | $\dfrac{\pi}{4}-\dfrac{2\sqrt{2}}{\Theta_S}$ | $\dfrac{\langle\Gamma\rangle}{\sqrt{2\omega}}\dfrac{1}{4\langle\varepsilon\rangle\sigma T_{V0}^3}$ |
| Volatile slab | $\dfrac{\omega\langle m_{V0}c_V\rangle}{\langle\varepsilon\rangle\sigma T_{V0}^3}$ | $1-\dfrac{\Theta_V}{32}$ | $\dfrac{4}{\Theta_V}\left(1-\dfrac{8}{\Theta_V^2}\right)$ | $\dfrac{\Theta_V}{4}$ | $\dfrac{\pi}{2}-\dfrac{4}{\Theta_V}$ | $\dfrac{\langle m_{V0}c_V\rangle}{4\langle\varepsilon\rangle\sigma T_{V0}^3}$ |
| Atmosphere | $\omega\dfrac{L}{f_V}\dfrac{dm_A}{dT_V}\dfrac{1}{\langle\varepsilon\rangle\sigma T_{V0}^3}$ | $1-\dfrac{\Theta_A}{32}$ | $\dfrac{4}{\Theta_A}\left(1-\dfrac{8}{\Theta_A^2}\right)$ | $\dfrac{\Theta_A}{4}$ | $\dfrac{\pi}{2}-\dfrac{4}{\Theta_A}$ | $\dfrac{L}{f_V}\dfrac{dm_A}{dT_V}\dfrac{1}{4\langle\varepsilon\rangle\sigma T_{V0}^3}$ |

### Table II: Example thermal parameters and time constants for Pluto's diurnal cycle

| | Assumptions | Thermal parameter | $\Delta T_V$ (K) | $p_{max}/p_{min}$ | $\tau$ (Earth days) |
|---|---|---|---|---|---|
| Unbuffered | Grundy & Fink (1996) map $A_V = 0.8$, $\varepsilon_V = 0.55$, $\Delta = 32.05$ AU, $\phi_0 = 44.25°$ $T_{V0} = 37.87$ K | n/a | 4.2 | 11.7 | n/a |
| Substrate | $\Gamma = 1.8\times10^4$ erg cm$^{-2}$ K$^{-1}$ s$^{-1/2}$ | $\Theta_S(T_{V0})=36$ | 0.41 | 1.27 | 6.4 |
| Volatile slab | $c_V = 1.3\times10^7$ erg g$^{-1}$ $m_{V0} = 9.5$ g cm$^{-2}$ | $\Theta_V(T_{V0})=830$ | 0.018 | 1.011 | 211 |
| Atmosphere | $L=2.532\times10^9$ erg g$^{-1}$ $f_V=0.53$ $g=61.9$ cm s$^{-2}$ $dp/dT = 10.36\,\mu$bar/K | $\Theta_A(T_{V0})=5411$ | 0.0025 | 1.0015 | 1375 |

Table III: Example thermal parameters and time constants for Pluto's seasonal cycle

|  | Assumptions | Thermal parameter | $\Delta T_V$ (K) | $p_{max}/p_{min}$ | $\tau$ (Earth years) |
|---|---|---|---|---|---|
| Unbuffered | "ice ball" $A_V = 0.8$, $\varepsilon_V = 0.55$, $\Delta = 29.7\text{-}49.1$ AU $F = 6.0$ erg cm$^{-2}$ s$^{-1}$ $T_{V0} = 35.83$ K | n/a | 8.1 | 213 | n/a |
| Substrate | $\Gamma = 2.9 \times 10^5$ erg cm$^{-2}$ K$^{-1}$ s$^{-1/2}$ | $\Theta_S(T_{V0}) = 5.8$ | 3.43 | 10.2 | 40 |
| Volatile slab | $c_V = 1.3 \times 10^7$ erg g$^{-1}$ $m_{V0} = 200$ g cm$^{-2}$ | $\Theta_V(T_{V0}) = 1.5$ | 3.17 | 8.6 | 14 |
| Atmosphere | $L = 2.536 \times 10^9$ erg g$^{-1}$ $f = 1$ $g = 61.9$ cm s$^{-2}$ $dp/dT = 3.14$ $\mu$bar/K | $\Theta_A(T_{V0}) = 0.074$ | 3.16 | 8.5 | 0.7 |

# Figures

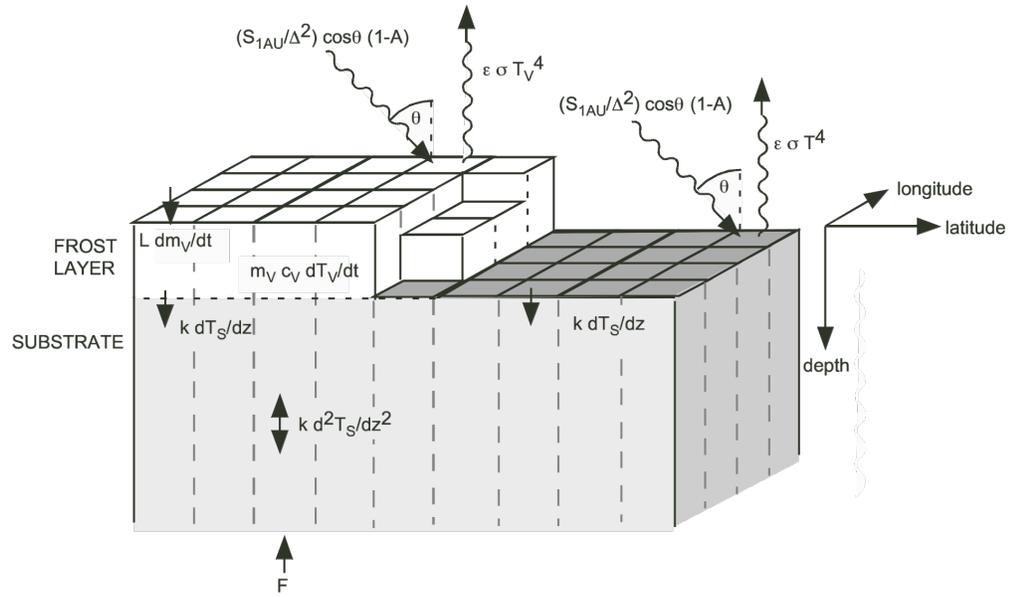

Fig 1.

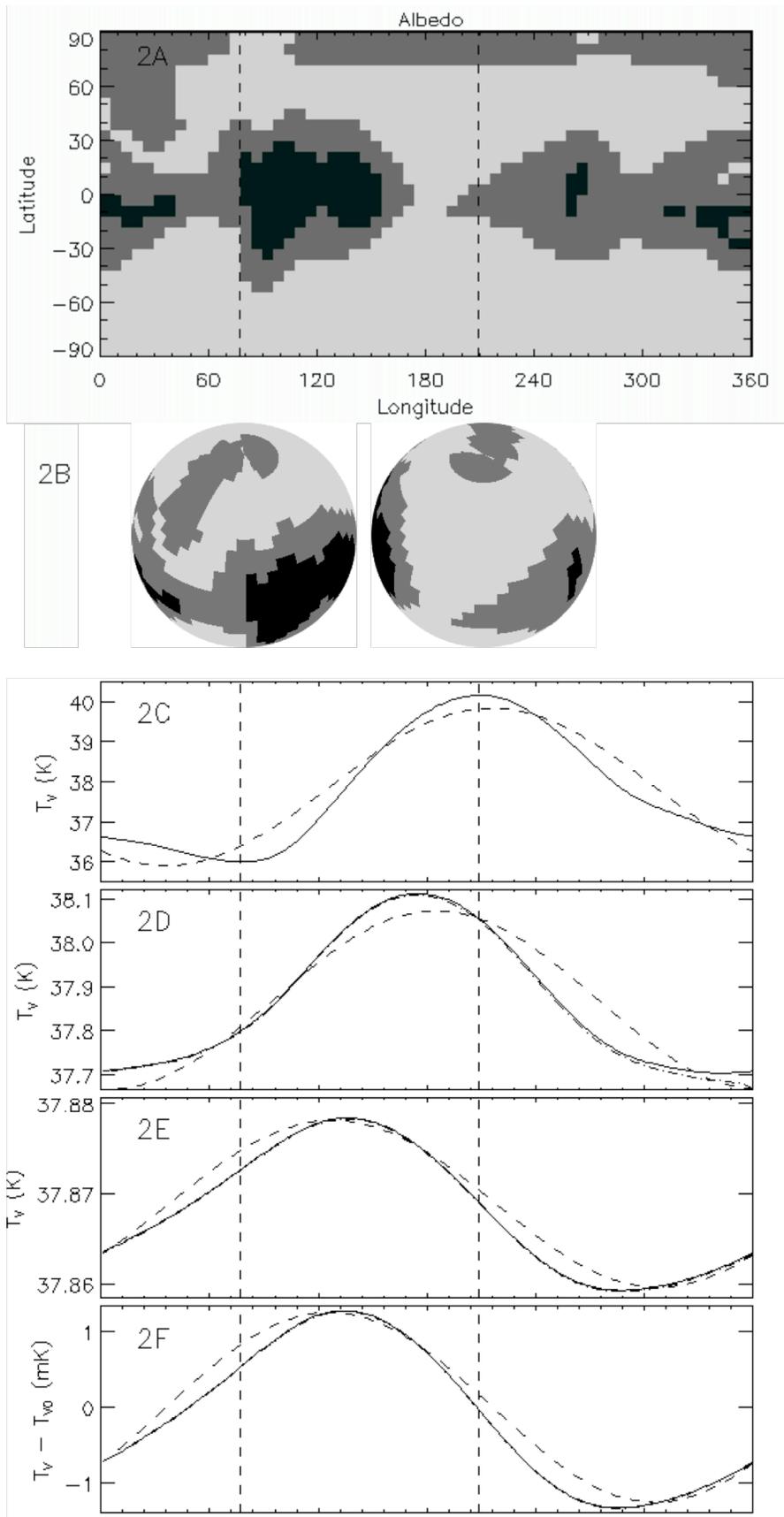

Fig 2.

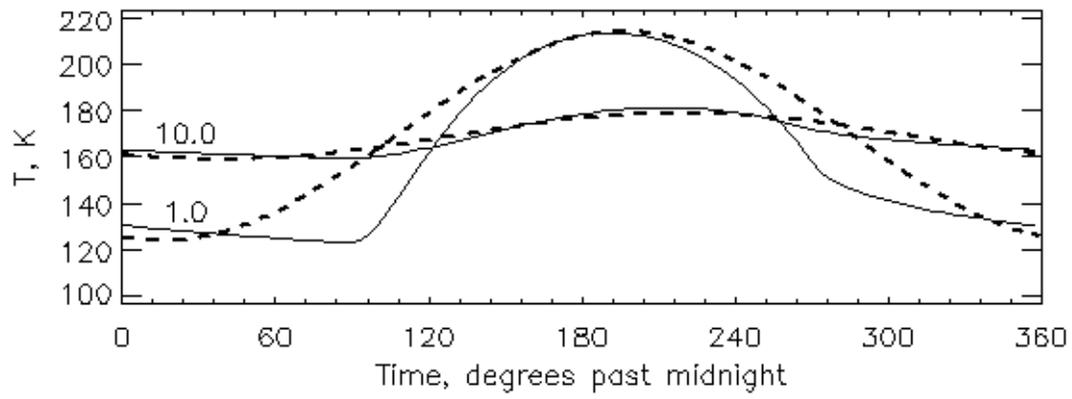

Fig. 3

## Figure Captions

Fig. 1. Schematic of the heat balance equation solved by the analytic model (based on Hansen and Paige, 1996). Locally, we balance incoming insolation, $S = (S_{1AU}/\Delta^2)\cos\theta(1\text{-}A)$, emitted thermal energy $\epsilon\sigma T^4$, and latent heat of sublimation or condensation, $L\ dm_V/dt$. Additionally we balance heat to and from the substrate, $k\ dT_S/dz$, the heat capacity of the ice slab, $m_V\ c_V\ dT_V/dt$, and heat flux at the lower boundary, $F$. All variables except $T_V$ are free to vary with latitude and longitude.

Fig. 2. Application to Pluto's diurnal rotation. 2A: Pluto's composition map, based on Grundy and Fink (1996), using their rotational-north-pole, sub-Charon prime meridian convention for latitude and longitude. The lightest terrains are $N_2$-rich, and participate in the global exchange of mass and energy. The mid-gray terrains are $CH_4$-rich, and the darkest terrains are tholins or $H_2O$. Also marked are the longitudes of minimum and maximum solar insolation (averaged over the frost), 77° and 209°, respectively. (2B) Pluto as seen from the sun in 2011, with a subsolar latitude of 44.25°. The two projections show Pluto when the $N_2$ ice sees the least insolation (left, 77°), and when it sees the most insolation (right, 209°). (2C) $N_2$-ice temperature for a surface in instantaneous equilibrium with the insolation. The solid line is the exact solution, and the dashed line is the first-order approximation to a sinusoid. (2D) $N_2$-ice temperatures, including the effects of a thermal wave within the substrate. The solid line is the numerical integration, and the dashed line is the analytic approximation presented here. The dot-dashed line most clearly seen near longitudes 300-360° plots the first period of the numerical solution, using the analytic solution as an initial condition, showing rapid convergence to the final numerical solution. In the rotational-north-pole, sub-Charon prime meridian convention, subsolar longitude decreases with time; a time lag translates to a maximum temperature that occurs at a smaller subsolar longitude than for 2C. (2E) $N_2$-ice temperatures, including the buffering needed to heat or cool an isothermal volatile slab. As in 2D, the solid line is the numerical integration, and the dashed line is the analytic approximation. In both 2E and 2F, convergence is very rapid, and the first period is indistinguishable from the final numerical solution. (2F) The difference between the time-varying $N_2$-ice temperature and its mean, when the atmospheric buffering is also included. As with 2D and 2E, the solid line is the numerical integration, and the dashed line is the analytic approximation. Note change in the $y$ axis from K to mK. Details on numeric integrations are described in the second paper in this series (Young 2012, in preparation), and follow a process similar to Hansen and Paige (1996) and Spencer and Moore (1992). In this example,

forward-stepping calculations were performed on a 6x6° spatial grid, with 2.5 layers per skin depth, and 240 points per period.

Fig. 3. Comparison of analytic (dashed) vs. numeric (solid) solutions for the surface temperature of frost-free regions, after Spencer et al. (1989). Two cases are plotted, one with $\Theta_S(T_{ss}) = 1$ ($\Theta_S(\hat{T}_{V0}) = 2.4$) and one with $\Theta_S(T_{ss}) = 10$ ($\Theta_S(\hat{T}_{V0}) = 24$). Both cases are for an equatorial location for an object at 3 AU, with an albedo of 0.05, an emissivity of 1.0, and a zero sub-solar latitude.